\documentclass[preprint]{elsarticle}
\usepackage{amssymb}
\usepackage{graphicx}

\journal{Physics Letters B}

\begin{document}

\begin{frontmatter}
\title{Reconstruction of Scalar Potentials in Induced Gravity and Cosmology}
\author{Alexander Y. Kamenshchik\corref{cor1}}
\address{Dipartimento di Fisica and INFN, Via Irnerio 46,40126 Bologna,
Italy\\
L.D. Landau Institute for Theoretical Physics of the Russian
Academy of Sciences, Kosygin str. 2, 119334 Moscow, Russia}\ead
{Alexander.Kamenshchik@bo.infn.it}
\author{Alessandro Tronconi}
\address{Dipartimento di Fisica, Via Irnerio 46,40126 Bologna,
Italy}\ead{Alessandro.Tronconi@bo.infn.it}
\author{Giovanni Venturi}
\address{Dipartimento di Fisica and INFN, Via Irnerio 46,40126 Bologna,
Italy}\ead{Giovanni.Venturi@bo.infn.it}
\cortext[]{Corresponding author}
\begin{abstract}
We develop a technique for the reconstruction of the potential for a scalar field in 
cosmological models based on induced gravity. The potentials reproducing cosmological 
evolutions driven by barotropic perfect fluids, a cosmological constant, a Chaplygin gas and a modified 
Chaplygin gas are constructed explicitly. 
\end{abstract}

\end{frontmatter}
\section{Introduction}
In the conventional formulation of General Relativity gravitation is described by adding to matter an Einstein-Hilbert gravitational action containing Newton's constant. An entirely different approach is that of induced gravity \cite{sakharov} wherein the gravitational constant and interaction arise as a quantum effect. In particular as a one-loop effect associated with the coupling of the curvature scalar to some hitherto unknown scalar field. Thus gravity itself would not be associated with ``fundamental physics'' but would be an ``emergent effect'' in which the conventional formulation is a low energy limit.\\
The application of induced gravity models to cosmology has been the subject of interest for several years \cite{Cooper:1982du}--\cite{CervantesCota:2010cb}. 
As is well-known scalar fields play an essential role in modern cosmology since they are possible candidates for the role of the inflaton field driving inflation in the early universe \cite{inflation} and of the dark energy substance \cite{dark} responsible for the present cosmic acceleration \cite{cosmic}. Thus, the technique of the reconstruction of the potentials of scalar fields reproducing a given cosmological evolution has attracted the attention of researchers for a long time
\cite{Starobinsky:1998fr}--\cite{Guo:2005ata}. The reconstruction of potentials for scalar fields non-minimally coupled to gravity was considered in \cite{Boisseau:2000pr}. The reconstruction of potentials for models with two scalar fields was studied in \cite{Andrianov:2007ua}, while a similar procedure for tachyon models was discussed in \cite{Padmanabhan:2002cp,Feinstein:2002aj,Gorini:2003wa}.

In the second section of the present paper we present a general technique for the reconstruction of scalar field potentials for a scalar field non-minimally coupled to gravity and then consider its version for the case of induced gravity \cite{induced}.
In the third section we explicitly construct the potentials which reproduce the cosmological evolutions driven by such perfect fluids as barotropic fluids with a fixed relation between the energy density and pressure, a cosmological constant, the Chaplygin gas \cite{we-Chap} and for the whole family of models, which is sometimes called ``modified Chaplygin gas'' \cite{modif}. By the cosmological evolution driven by some perfect fluid characterized by certain energy density $\varepsilon$  and pressure $p$ we mean the evolution of the flat Friedmann universe with the scale factor $a$ described by the Hubble parameter $h=\dot{a}/a$, where a ``dot'' represents the time derivative.  The energy density and the pressure are related to the Hubble parameter by the Friedmann equations, which we write in the form
\begin{equation}
\varepsilon=h^2,
\label{fr1}
\end{equation}
and 
\begin{equation}
p = -h^2-\frac{2}{3}\dot{h},
\label{fr2}
\end{equation}
where a convenient choice of Newton's constant $G$, $8\pi G/3 = 1$, has been made.

We also consider separately the cases of the conformal coupling ($-1/6$) and the coupling with the coefficient equal to $-1/4$ because for these cases it is possible to obtain some general expressions for the potentials. 

Section 4 contains some concluding remarks. 

\section{The basic equations}
Let us consider the action
\begin{equation}
S = \int dx \sqrt{-g} \left(U(\sigma) R - \frac12 g^{\mu\nu}\sigma_{,\mu}\sigma_{\,\nu} + V(\sigma)\right).
\label{action}
\end{equation}
In a Friedmann-Robertson-Walker flat spacetime with metric 
\begin{equation}
ds^2 = N^2(t)dt^2 - a^2(t)(dx^2+dy^2+dz^2),
\label{metric}
\end{equation}
the scalar curvature is 
\begin{equation}
R = -6\left(\frac{\ddot{a}}{aN^2} - \frac{\dot{N}\dot{a}}{aN^3}+\frac{\dot{a}^2}{a^2N^2}\right).
\label{curvature}
\end{equation}
and 
\begin{equation}
\sqrt{-g} = Na^3.
\label{determ}
\end{equation}
The Lagrangian in this minisuperspace is 
\begin{equation}
L = -6U\left(\frac{\ddot{a}a^2}{N^2} - \frac{\dot{N}\dot{a}a^2}{N^2} + \frac{\dot{a}^2a}{N}\right) -\frac{a^3\dot{\sigma}^2}{2N}
+ Na^3V.
\label{Lagrange}
\end{equation}
After integration by parts 
\begin{equation}
L = \frac{6U\dot{a}^2a}{N}-\frac{a^3\dot{\sigma}^2}{2N}+6\frac{dU}{d\sigma}\frac{\dot{\sigma}\dot{a}a^2}{N} + Na^3V.
\label{Lagrange1}
\end{equation}
The variation of the Lagrangian (\ref{Lagrange1}) with respect to the lapse function $N$ gives 
\begin{equation}
-\frac{6U\dot{a}^2a}{N^2} + \frac{a^3\dot{\sigma}^2}{2N^2} -6\frac{dU}{d\sigma}\frac{\dot{\sigma}\dot{a}a^2}{N^2} 
+ a^3V = 0. 
\label{Fried}
\end{equation}
Fixing as a gauge condition $N = 1$ and dividing Eq. (\ref{Fried}) by $a^3$ we obtain the generalized Friedmann equation
\begin{equation}
h^2 = \frac{\dot{\sigma}^2}{12U}+\frac{V}{6U} - \frac1U\frac{dU}{d\sigma}h\dot{\sigma}.
\label{Fried1}
\end{equation}
The variation of the Lagrangian (\ref{Lagrange1}) with respect to $a$ gives, after division by $a^2$, the second 
Friedmann equation
\begin{equation}
12U\dot{h} + 18Uh^2 + 12\frac{dU}{d\sigma}\dot{\sigma}h + 6\frac{dU}{d\sigma}\ddot{\sigma}+6\frac{d^2U}{d\sigma^2}\dot{\sigma}^2
+\frac32\dot{\sigma}^2-3V = 0.
\label{Fried2}
\end{equation}
Finally, the variation of the Lagrangian (\ref{Lagrange1}) with respect to $\sigma$ gives the Klein-Gordon equation
\begin{equation}
\ddot{\sigma} + 3h\dot{\sigma} + \frac{dV}{d\sigma} - 6\frac{dU}{d\sigma}(\dot{h}+2h^2) = 0.
\label{KG}
\end{equation}

Let us now suppose that the function $U(\sigma)$ is fixed.
Then suppose that we have some prescribed cosmological evolution given by the Hubble parameter $h$ as a function of the cosmic time $t$ or of the cosmological radius $a$. If we know the dependence of the scalar field $\sigma$ on the variables $t$ 
or $a$ and if we can invert the corresponding function, we are able to find an explicit form for the potential compatible with 
the given evolution, provided some special initial conditions are chosen. This form of the potential can be obtained from the 
equation 
\begin{equation}
V = 6Uh^2-\frac{\dot{\sigma}^2}{2}+6\frac{dU}{d\sigma}\dot{\sigma}h,
\label{poten}
\end{equation}
which is a consequence of Eq. (\ref{Fried1}). In what follows it will be more convenient to consider all the functions as functions of $a$ and not of $t$. Correspondingly, Eq. (\ref{poten}) can rewritten as
\begin{equation}
V = h^2\left(6U-\frac{\sigma'^2a^2}{2}+6\frac{dU}{d\sigma}\sigma'a\right),
\label{poten1} 
\end{equation}
where a ``prime'' denotes the derivative with respect to $a$.
We would now like to eliminate the potential $V$ from the Klein-Gordon equation (\ref{KG}).
We can find the derivative $dV/d\sigma$, by using Eq. (\ref{poten})
\begin{eqnarray}
&&\frac{dV}{d\sigma} = \frac{\dot{V}}{\dot{\sigma}}
=6\frac{dU}{d\sigma}h^2+\frac{12U\dot{h}h}{\dot{\sigma}}-\ddot{\sigma}+6\frac{d^2U}{d\sigma^2}\dot{\sigma}h 
+ 6\frac{dU}{d\sigma}\frac{\ddot{\sigma}h}{\dot{\sigma}}+6\frac{dU}{d\sigma}\dot{h}.
\label{poten2}
\end{eqnarray}
On substituting the expression (\ref{poten2}) into the Klein-Gordon equation (\ref{KG}) we obtain 
\begin{equation}
3\dot{\sigma}^2-6\frac{dU}{d\sigma}\dot{\sigma}h+12U\dot{h} + 6\frac{d^2U}{d\sigma^2}\dot{\sigma}^2+6\frac{dU}{d\sigma}\ddot{\sigma} = 0.
\label{master}
\end{equation}
For the case of a minimal coupling to gravity ($U = const$) Eq. (\ref{master}) reduces to 
\begin{equation}
\dot{\sigma}^2+ 4U\dot{h} = 0
\label{minim}
\end{equation}
or, if one wishes to work with the dependence of $\sigma$ on $a$, it becomes
\begin{equation}
\sigma'^2 + 4U\frac{h'}{ha} = 0.
\label{minim1}
\end{equation}
The reconstruction procedure for potentials of minimally coupled fields is based on the use of formulae (\ref{minim}) 
or (\ref{minim1}). 
For the case of a nonminimal coupling it is more convenient to use the dependence of the scalar field on the cosmological radius 
$a$. Correspondingly Eq. (\ref{master}) should be rewritten as
\begin{equation}
\sigma''+ \sigma'\frac{h'}{h} 
+ \sigma'^2\left(\frac{1+2\frac{d^2U}{d\sigma^2}}{2\frac{dU}{d\sigma}}\right) 
  + \frac{2Uh'}{\frac{dU}{d\sigma}ha} = 0.
\label{master1}
\end{equation}
For the case of induced gravity \cite{induced}
\begin{equation}
U(\sigma) = \frac12\gamma\sigma^2,
\label{ind}
\end{equation}
Eq. (\ref{master1}) acquires a simpler form
\begin{equation}
\sigma''+ \sigma'\frac{h'}{h} + \sigma'^2\frac{2\gamma + 1}{2\gamma \sigma} + \sigma\frac{h'}{ha} = 0.
\label{master2}
\end{equation}
On now introducing the variable 
\begin{equation}
x \equiv \frac{\sigma'}{\sigma},
\label{x-def}
\end{equation}
we can rewrite Eq. (\ref{master2}) in the following form
\begin{equation}
x' + x\frac{h'}{h} + x^2\frac{4\gamma + 1}{2\gamma} + \frac{h'}{ha} = 0.
\label{master3}
\end{equation}
On introducing a new function $f$ such that 
\begin{equation}
x = \frac{2\gamma}{4\gamma + 1}\frac{f'}{f}
\label{f-def}
\end{equation}
we obtain the following linear second order differential equation for $f$:
\begin{equation}
f'' + f'\frac{h'}{h} + f\frac{1+4\gamma}{2\gamma}\frac{h'}{ha} = 0.
\label{master4}
\end{equation}
It is easy to see, on comparing Eqs. (\ref{x-def}) and (\ref{f-def}), that 
\begin{equation}
\sigma = (f)^{\frac{2\gamma}{4\gamma + 1}}.
\label{sigma-f}
\end{equation}
Finally, formula (\ref{poten1}) for the potential $V$ in the case of the induced gravity becomes
\begin{equation}
V = h^2\left(3\gamma\sigma^2 - \frac12\sigma'^2a^2 +6\gamma \sigma\sigma'a\right).
\label{poten3}
\end{equation}

All the considerations following Eq. (\ref{poten2}) are valid providing the time derivative of the 
scalar field $\sigma$ is different from zero. The case of the constant field $\sigma$ must be treated separately. 
Let us first of all notice that, if $\sigma$ is constant, then the functions $U$ and $V$ also do not depend on time and it follows immediately from the Friedmann equation (\ref{Fried1}) that the Hubble parameter $h$ is constant too. This corresponds to a cosmological evolution driven by the cosmological constant. Eq. (\ref{Fried1}) can now be rewritten as 
\begin{equation}
V = 6Uh_0^2.
\label{pot-const}
\end{equation}
Then, on substituting $\dot{\sigma} = 0$ and $\dot{h} = 0$ into the Klein-Gordon equation (\ref{KG}) we obtain
\begin{equation}
\frac{dV}{d\sigma} = 12\frac{dU}{d\sigma}h_0^2.
\label{KG-const}
\end{equation}
Dividing Eq. (\ref{KG-const}) by Eq. (\ref{pot-const}) we obtain
\begin{equation}
\frac1V\frac{dV}{d\sigma} = 2\frac1U\frac{dU}{d\sigma},
\label{KG-Fr}
\end{equation}
which gives 
\begin{equation}
V = V_0 U^2,
\label{pot-const1}
\end{equation}
where $V_0$ is an arbitrary constant. For the case of induced gravity (\ref{ind}) we obtain from Eq. (\ref{pot-const1}) 
a quartic potential
\begin{equation}
V = \lambda\sigma^4.
\label{quartic}
\end{equation}


\section{Examples}
\subsection{Barotropic fluid}
For a barotropic fluid with the equation of state
\begin{equation}
p = w\varepsilon,
\label{bar}
\end{equation}
the Hubble parameter behaves as
\begin{equation}
h(a) = \frac{h_0}{a^{\frac32(w+1)}}
\label{bar1}
\end{equation}
and 
\begin{equation}
\frac{h'}{h} = -\frac{3(w+1)}{2a}. 
\label{bar2}
\end{equation}
The basic equation (\ref{master4}) is now
\begin{equation}
f''+\frac{\alpha}{a}f' + \frac{\alpha\beta}{a^2} f = 0,
\label{master-bar} 
\end{equation}
 where we have introduced the following notation
\begin{equation}
\alpha \equiv -\frac32(w+1),\ \ \beta \equiv \frac{1+4\gamma}{2\gamma}.
\label{alphabeta}
\end{equation}
The general solution of Eq. (\ref{master-bar}) is 
\begin{equation}
f(a) = c_1a^{p_1} + c_2a^{p_2},
\label{bar-sol}
\end{equation}
where $c_1$ and $c_2$ are arbitrary constants, and the exponents $p_1$ and $p_2$ are
\begin{equation}
p_{1,2} = \frac{1-\alpha}{2} \pm \sqrt{\left(\frac{1-\alpha}{2}\right)^2-\alpha\beta}.
\label{exponents}
\end{equation}
We shall always choose one of the constants $c_1$ or $c_2$ equal to zero. Only in this case one can hope 
to invert the function (\ref{bar-sol}).
Then, up to a constant
\begin{equation}
a = f^{\frac{1}{p_{1,2}}} 
\label{a-f}
\end{equation}
and from Eq. (\ref{sigma-f}) it follows that 
\begin{equation}
\sigma = a^{\frac{p_{1,2}}{\beta }}
\label{sigma-a}
\end{equation}
 and inversely
\begin{equation}
a = \sigma^{\frac{\beta}{p_{1,2}}}.
\label{a-sigma}
\end{equation}
Another useful formula is 
\begin{equation}
\sigma'a = \frac{p_{1,2}}{\beta}\sigma.
\label{a-sigma1}
\end{equation}
On substituting the formulae (\ref{sigma-a}) and (\ref{a-sigma1}) into the potential 
(\ref{poten3}) we obtain
\begin{equation}
V_{1,2} = V_0 \left[\frac{6\gamma\beta^2 +\alpha\beta+(\alpha-1+12\gamma\beta)  
 p_{1,2}}{2\beta^2}\right]\sigma^{\frac{2(\alpha\beta+p_{1,2})}{p_{1,2}}},
\label{poten4}
\end{equation}
where $V_0$ is an arbitrary constant.

We can now consider the limit $w = -1$ (cosmological constant) for the solutions (\ref{poten4}). 
In this case $p_1 = 1$ and $p_2 = 0$. If we choose $c_1 \neq 0, c_2 = 0$ then the formula (\ref{poten4}) gives us 
the quadratic potential
\begin{equation}
V_1 = V_0\left(\frac{6\gamma\beta^2-1+12\gamma\beta}{2\beta^2}\right)\sigma^2.
\label{quadratic}
\end{equation}
On choosing $c_1 =1, c_2  \neq 0$, we should consider the limit of the exponent in the formula (\ref{poten4}) when 
$w, \alpha$ and $p_2$ tend to zero. This limit is equal to 4 and we obtain the quartic potential 
\begin{equation}
V_2 = 3\gamma V_0\sigma^4,
\label{quartic1}
\end{equation}
which coincides with that already obtained at the end of the second section (see Eq. (\ref{quartic})). 
The general form of the potential for the case of a cosmological constant will be obtained in the next subsection. 

\subsection{Cosmological constant}
It is best to treat the case of a cosmological constant separately, because in this case 
it is possible to invert the general solution for the function $\sigma(a)$. 
In this case the Hubble parameter does not depend on $a$, hence $h' = 0$, and Eq. (\ref{master4}) has the form 
$f'' = 0$ and its general solution is 
\begin{equation}
f(a) = c_1 + c_2a.
\label{cosm-const} 
\end{equation}
Correspondingly from Eq. (\ref{sigma-f})  
\begin{equation}
\sigma(a) = (c_1 + c_2a)^{\frac{1}{\beta}} 
\label{sigma-cosm}
\end{equation}
and, if $c_2 \neq 0$,
\begin{equation}
a = \frac{\sigma^{\beta}-c_1}{c_2}.
\label{a-cosm}
\end{equation}
The case $c_2 = 0$ should be treated separately. In this case 
$\sigma$ does not depend on $a$ and we already know that the corresponding potential  
has the quartic form (see Eqs. (\ref{quartic}) and (\ref{quartic1})).

If $c_2 \neq 0$, using the formulae (\ref{sigma-cosm}) and (\ref{a-cosm}) we obtain
\begin{equation}
\sigma' a = \frac{\sigma}{\beta}\left(1 - c_1\sigma^{-\beta}\right).
\label{sigma-prime}
\end{equation}
On substituting Eq. (\ref{sigma-prime}) into Eq. (\ref{poten4}) we obtain 
\begin{equation}
V =h_0^2\frac{\gamma^{2}(1+6\gamma)}{(4\gamma+1)^2}\left[\frac{(16\gamma+3)}{\gamma}\sigma^2-\frac{2\,c_1^2}{(1+6\gamma)}\sigma^{-\frac{2\gamma+1}{\gamma}}
-8\,c_1\sigma^{-\frac{1}{2\gamma}}\right].
\label{poten-cosm1}
\end{equation}

\subsection{Conformal coupling $\gamma = -\frac16$}
In this case $\beta = -1$. For the case of the perfect fluid $p_1 = -\alpha$ and $p_2 = 1$. 
For the first case the potential is equal to 
\begin{equation}
V_1 = -\frac{V_0(\alpha+1)^2}{2}\sigma^{4},
\label{pot-conf}
\end{equation}
while the potential corresponding to $p_2 = 1$ vanishes.
For a cosmological constant, the potential corresponding to the case $c_2 = 0$ 
will have the form
\begin{equation}
V = -\frac{V_0\sigma^{4}}{2},
\label{pot-conf1}
\end{equation}
while the potential given by formula (\ref{poten-cosm1}) will have the form 
\begin{equation}
V = -\frac{c_1^2h_0^2\sigma^4}{2}.
\label{pot-conf2}
\end{equation}
Generally equation (\ref{master4}) for $f(a)$ in this case has a simple form
\begin{equation}
f''+\frac{h'}{h}f' -\frac{h'}{ha}f = 0.
\label{master-conf}
\end{equation}
One of its solutions 
\begin{equation}
f_1(a) = a
\label{sol-conf}
\end{equation}
does not depend on $h(a)$ and gives a potential equal to zero. The second solution 
\begin{equation}
f_2 = a\int\frac{da}{a^2h}
\end{equation}
cannot be integrated explicitly for an arbitrary $h$. 
\subsection{The case $\gamma = -\frac14$}
This case should be treated separately, because $\beta = 0$ and some of the formulae used in the subsections 3.1 and 3.2 are meaningless. The  equation (\ref{master3}) for the variable $x$ is now 
\begin{equation}
x' + x\frac{h'}{h} + \frac{h'}{ha} = 0.
\label{x-special}
\end{equation}
Its general solution is 
\begin{equation}
x = \frac{c_3}{h} - \frac{1}{h}\int\frac{h'}{a}da.
\label{x-special1} 
\end{equation}
We shall write down this solution for some chosen forms of $h(a)$. For the case of the cosmological constant $h'= 0$ 
and 
\begin{equation}
x = c_4,
\label{x-spectial2}
\end{equation}
where $c_4$ is an arbitrary constant, from Eq. (\ref{x-def}) we have immediately 
\begin{equation}
\sigma = \sigma_0e^{c_4a},
\label{sigma-special}
\end{equation}
where $\sigma_0$ is a constant.
Then 
\begin{equation}
a = \frac{1}{c_4}\ln \frac{\sigma}{\sigma_0}.
\label{a-special}
\end{equation}
On substituting Eqs. (\ref{sigma-special}) and (\ref{a-special}) into Eq. (\ref{poten3}) 
we obtain
\begin{equation}
V = h_0^2 \left(-\frac34\sigma^2 -\frac12\sigma^2\ln^2\frac{\sigma}{\sigma_0} -\frac32\sigma^2\ln\frac{\sigma}{\sigma_0}\right).
\label{pot-special}
\end{equation}
At first glance, it seems that there is no connection between the last formula and the general formula of the potential 
for the case of the cosmological constant (see Eq. (\ref{poten-cosm1})). Indeed, the formula (\ref{poten-cosm1}) seems to be 
inapplicable to the case $\gamma = -\frac14$ because of the presence of factor $(4\gamma+1)^2$ in the denominator on the 
right-hand side of Eq. (\ref{poten-cosm1}). This is not correct. In fact, by choosing the arbitrary 
constant $c_1$ such as 
\begin{equation}
c_1 = \sigma_0^{2+\frac{1}{2\gamma}},
\end{equation}
and substituting it into Eq. (\ref{poten-cosm1}), in the limit for $\gamma \rightarrow -\frac{1}{4}$, the singularity in the denominator is cancelled by the corresponding zero of the second order in the numerator and one is left with 
an expression for the potential which coincides with that given by Eq. (\ref{pot-special}).  

For the barotropic fluid, using formula (\ref{bar1}), one finds from Eq. (\ref{x-special1}) a general solution for $x$ given by
\begin{equation}
x = c_5\, a^{\frac32(w+1)}-\frac{3w+3}{3w+5}\,a^{-1}.
\label{x-spec1}
\end{equation}
We shall choose $c_5 = 0$, because only in this case one can find $a$ as an explicit function of $\sigma$.
Then 
\begin{equation}
\sigma = \sigma_0a^{-\frac{3w+3}{3w+5}} 
\label{sigma-sp}
\end{equation}
and 
\begin{equation}
a = \left(\frac{\sigma}{\sigma_0}\right)^{-\frac{3w+5}{3w+3}}.
\label{a-sp}
\end{equation}
Finally, the potential for this case is
\begin{equation}
V = \frac{V_0(3w^2+6w-1)}{(3w+5)^2}\sigma^{3w+7}.
\label{pot-sp}
\end{equation}
\subsection{Chaplygin gas}
The cosmological model of the Chaplygin gas \cite{we-Chap}
has acquired a certain popularity being, perhaps, a simple unified model 
of dark energy and dark matter \cite{we-Chap,Bilic,we-Chap1} 
The equation of state for the Chaplygin gas is 
\begin{equation}
p = -\frac{A}{\varepsilon}.
\label{Chap}
\end{equation}
Correspondingly the Hubble parameter is 
\begin{equation}
h(a) = \left(A + \frac{B}{a^6}\right)^{\frac14}.
\label{Hubble-Chap}
\end{equation}
The equation (master) for $f$ is now 
\begin{equation}
f''-\frac{3B}{2a(Aa^6+B)}f' - \frac{3B\beta}{2a^2(Aa^6+B)}f = 0.
\label{master-Chap}
\end{equation}
The general solution to this equation is 
\begin{eqnarray}
f(a)&=& c_1a^{\frac14(5-\delta)}\ _{2}F_1\left(\frac{1}{24}-\frac{\delta}{24};
\frac{5}{24}-\frac{\delta}{24};1-\frac{\delta}{12};-z\right)\nonumber \\
&&+c_2a^{\frac14(5+\delta)}\ _2F_1\left(\frac{1}{24}+\frac{\delta}{24};
\frac{5}{24}+\frac{\delta}{24};1+\frac{\delta}{12};-z\right),
\label{hypergeom}
\end{eqnarray}
where $\delta\equiv \sqrt{25+24\beta}$ and $z \equiv \frac{A}{B}a^6$.
In principle, to find a solution invertible with respect to $a$ one can put $c_2 =0$ and then search for values of the parameter $\beta$ (and hence $\gamma$) such that the hypergeometric series terminates. 
However, we shall choose a simpler approach. In any case, we shall assume for the invertible function $f$ the form
\begin{equation}
f(z) = \frac{r_0 + r_1z}{s_0+s_1z}.
\label{ansatz}
\end{equation}
On substituting this ansatz into Eq. (\ref{master-Chap}) we find that the only acceptable values of the parameters $\beta, r_0,r_1,s_0,s_1$  are 
the following :
\begin{equation}
\beta = 34,
\label{beta34}
\end{equation}
\begin{equation}
s_0=0,\ r_1 = \frac{14}{17}r_0.
\label{sr}
\end{equation}
From (\ref{beta34}) one has that 
\begin{equation}
\gamma = \frac{1}{64},
\label{gamma64}
\end{equation}
while from Eq. (\ref{sr}) one has that 
\begin{equation}
\frac{1}{z} = \frac{f}{f_0} - \frac{14}{17},
\label{z-f}
\end{equation}
where $f_0$ is an arbitrary constant. 
Using relation (\ref{sigma-f}) 
we obtain
\begin{equation}
f =f_0 \left(\frac{\sigma}{\sigma_0}\right)^{34}.
\label{f-sigma-Chap}
\end{equation}
We shall also need the relation
\begin{equation}
\sigma' a = \frac{3}{17}\frac{\sigma}{\sigma_0}\left[\frac{14}{17}\left(\frac{\sigma}{\sigma_0}\right)^{-34}-1\right],
\label{sigma-a-Chap}
\end{equation}
where $\sigma_0$ is an arbitrary constant. The Hubble parameter is given 
by 
\begin{equation}
h^2 = \sqrt{A}\sqrt{\frac{3}{17}+\left(\frac{\sigma}{\sigma_0}\right)^{34}}.
\label{Hubble-Chap1}
\end{equation}
On substituting the formulae (\ref{gamma64})--(\ref{Hubble-Chap1}) into Eq. (\ref{poten3}) 
we obtain
\begin{eqnarray}
V &=& \frac{3}{2}\sqrt{A}\sigma^2\sqrt{\frac{3}{17}+\left(\frac{\sigma}{\sigma_0}\right)^{34}}
\left\{\frac{1}{32} 
-\frac{3}{289}\left[\frac{14}{17}\left(\frac{\sigma}{\sigma_0}\right)^{-34}-1 \right]^2\right.\nonumber \\
&&\left.+
\frac{3}{272}\left[\frac{14}{17}\left(\frac{\sigma}{\sigma_0}\right)^{-34}-1 \right]\right\}.
\label{pot-Chap}
\end{eqnarray}
\subsection{``Modified'' Chaplygin gas}
It is also possible consider the case of the ``modified'' Chaplygin gas, i.e. the perfect fluid
for which the dependence of the Hubble parameter on the cosmological radius is given by the formula:
\begin{equation}
h(a) = \left(A+\frac{B}{a^{\rho}}\right)^{\tau}.
\label{modified}
\end{equation}
Such a dependence of the Hubble parameter on the cosmological radius corresponds to a 
perfect fluid with the equation of state 
\begin{equation}
p = \left(\frac{2\rho\tau}{3}-1\right)\varepsilon -\frac{2\rho\tau}{3}A\varepsilon^{\frac{2\tau-1}{2\tau}}.
\label{modified1}
\end{equation}
The term ``modified Chaplygin gas''  in connection with this kind of equation of state was used in papers 
\cite{modif}. Examples of such an equation of state were also considered in \cite{Barrow} and \cite{we-Chap0}.
The case $\tau = \frac12$ corresponds to a mixture of two fluids: that of a cosmological constant, $p=-\epsilon$, and that of a fluid with the equation of state $p = \left(\frac{\rho}{3}-1\right)\varepsilon$. If the constants $\tau$ and $\rho$ are connected 
by the relation $\tau = \frac{3}{2\rho}$ one has a generalized Chaplygin gas \cite{we-Chap,Bento} while the case 
$\tau = \frac14, \rho =6$ describes the Chaplygin gas.

The master equation for the function $f$ is now  
\begin{equation}
f''-\frac{\rho\tau B}{a(Aa^{\rho} + B)}f'-\frac{\beta\rho\tau B}{a^2(Aa^{\rho} + B)}f = 0.
\label{master-mod}
\end{equation}
On using the same ansatz as for the case of the Chaplygin gas (\ref{ansatz}) with $z \equiv Aa^{\rho}/B$ we obtain analogous relations for the parameters:
\begin{equation}
s_0 = 0,\; r_1 = \frac{(\rho+1)r_0}{\rho+1+\rho\tau}
\label{modified2}
\end{equation}
and 
\begin{equation}
\beta = \frac{\rho+1+\rho\tau}{\tau},\ \gamma = \frac{\tau}{2(\rho+1+\rho\tau-2\tau)}.
\label{modified3}
\end{equation}
Correspondingly
\begin{equation}
\frac{1}{z} + \frac{\rho+1}{\rho+1+\rho\tau} = \left(\frac{\sigma}{\sigma_0}\right)^{\frac{\rho+1+\rho\tau}{\tau}}.
\label{modified4}
\end{equation}
Then
\begin{equation}
\sigma'a = \frac{\rho\tau}{\rho+1+\rho\tau}\sigma\left[ \frac{(\rho+1)}{(\rho+1+\rho\tau)}
\left(\frac{\sigma}{\sigma_0}\right)^{-\frac{\rho+1+\rho\tau}{\tau}}-1\right].
\label{modified5}
\end{equation}
The explicit expression for the potential is cumbersome:
\begin{eqnarray}
&&V = A^{2\tau}\sigma^{2}\left[\frac{\rho\tau}{\rho+1+\rho\tau} + \left(\frac{\sigma}{\sigma_0}\right)^{\frac{\rho+1+\rho\tau}{\tau}}\right]^{2\tau}
\nonumber \\
&&\times\left\{\frac{3\tau}{2(\rho+1+\rho\tau-2\tau)} - \frac12\frac{(\rho\tau)^{2}}{(\rho+1+\rho\tau)^{2}}\left[ \frac{(\rho+1)}{(\rho+1+\rho\tau)}
\left(\frac{\sigma}{\sigma_0}\right)^{-\frac{\rho+1+\rho\tau}{\tau}}-1\right]^2\right.\nonumber \\
&&\left.+\frac{6\rho\tau^{2}}{2(\rho+1+\rho\tau-2\tau)(\rho+1+\rho\tau)}\left[ \frac{(\rho+1)}{(\rho+1+\rho\tau)}
\left(\frac{\sigma}{\sigma_0}\right)^{-\frac{\rho+1+\rho\tau}{\tau}}-1\right]\right\}.
\label{poten-modified}
\end{eqnarray}
For the case of the modified Chaplygin gas an additional explicit solution also exists. 
If the parameters $\rho$ and $\tau$ are connected by the relation 
\begin{equation}
\tau = \frac{\rho^2+2\rho-1}{\rho}
\label{rho-tau}
\end{equation}
and if 
\begin{equation}
\beta = -\frac{\rho^2(\rho+1)}{\rho^2+2\rho-1},\; \gamma = \frac{\rho^2+2\rho-1}{2(-\rho^3-3\rho^2-4\rho+2)},
\label{beta-mod}
\end{equation}
then there is a solution (\ref{ansatz}) with the coefficients
\begin{equation}
r_0 = 0,\; s_1 = -\frac{s_0}{\rho}.
\label{mod-except} 
\end{equation}
Correspondingly,
\begin{equation}
\frac{1}{z} -\frac{1}{\rho} = \left(\frac{\sigma}{\sigma_0}\right)^{\frac{\rho^2(\rho+1)}{\rho^2+2\rho-1}}.
\label{z-mod}
\end{equation}
and 
\begin{equation}
\sigma'a =\frac{\rho^2(\rho+1)}{\rho^2+2\rho-1}\sigma\left[ \left(\frac{\sigma}{\sigma_0}\right)^{-\frac{\rho^{2}\left(\rho+1\right)}{\rho^2+2\rho-1}}
+\rho\right].
\label{mod-except1}
\end{equation}
We can now write down the explicit expression for the potential
\begin{eqnarray}
V&=&A^{\frac{2(\rho^2+2\rho-1)}{\rho}}\sigma^{2}\left[\frac{\rho+1}{\rho}+\left(\frac{\sigma}{\sigma_0}\right)^{\frac{\rho^2(\rho+1)}{\rho^2+2\rho-1}}\right]^{\frac{2(\rho^2+2\rho-1)}{\rho}}\nonumber \\
&&\times\left\{\frac{3(\rho^2+2\rho-1)}{2(-\rho^3-3\rho^2-4\rho+2)}\right.
-\frac12\frac{\rho^4(\rho+1)^{2}}{(\rho^2+2\rho-1)^{2}}\left[ \left(\frac{\sigma}{\sigma_0}\right)^{-\frac{\rho^{2}\left(\rho+1\right)}{\rho^2+2\rho-1}}
+\rho\right]^2
\nonumber \\
&&\left.+\frac{3\rho^2(\rho+1)}{-\rho^3-3\rho^2-4\rho+2}
\left[ \left(\frac{\sigma}{\sigma_0}\right)^{-\frac{\rho^{2}\left(\rho+1\right)}{\rho^2+2\rho-1}}
+\rho\right]\right\}.
\label{poten-mod-except}
\end{eqnarray}
\section{Conclusion}
In this letter we have extended the procedure of the reconstruction of the potentials for a scalar field to 
the case of induced gravity. We have obtained the explicit forms of potentials reproducing the dynamics of 
a flat Friedmann-Robertson-Walker universe, driven by barotropic perfect fluids, by a Chaplygin gas and by a modified 
Chaplygin gas. The case of a cosmological constant was considered separately, because in this case it was possible 
to construct a general formula for a family of potentials. We have also considered the potentials for some particular 
values of the coupling parameter $\gamma$, when rather general potentials can be constructed as well.
These values are $\gamma = -\frac16$, which is the well-known case of a conformal coupling and the case $\gamma = -\frac14$, which, to our knowledge, was not considered before.
\section*{Acknowledgements}
A.K. was partially supported by the RFBR  grant No 11-02-00643.


\begin{thebibliography}{99}
 \bibitem{sakharov}
A. D. Sakharov, Dokl. Akad. Nauk. SSSR 117 (1967)  70; 
[Sov. Phys. Dokl. 12 (1967) 1040].
\bibitem{Cooper:1982du}
  F.~Cooper and G.~Venturi,
  Phys.\ Rev.\  D 24 (1981) 3338.
\bibitem{induced}
 F. Finelli, A. Tronconi and G. Venturi,
  Phys. Lett.  B 659 (2008) 466.
\bibitem{Cerioni:2009kn}
  A.~Cerioni, F.~Finelli, A.~Tronconi and G.~Venturi,
  Phys.\ Lett.\  B  681 (2009) 383.
\bibitem{Cerioni:2010ke}
  A.~Cerioni, F.~Finelli, A.~Tronconi and G.~Venturi,
  Phys.\ Rev.\  D  81 (2010) 123505.
\bibitem{CervantesCota:2010cb}
  J.~L.~Cervantes-Cota, R.~de Putter and E.~V.~Linder,
  JCAP {\bf 1012} (2010) 019
\bibitem{inflation}
A.~A.~Starobinsky,
Lect. Notes in Phys. 246 (1986) 107;
 A. D. Linde, Particle Physics and Inflationary Cosmology, Chur, Switzerland: Harwood, 1990.
\bibitem{dark}
V. Sahni and A.A. Starobinsky, Int. J. Mod. Phys. D  9 (2000) 373;
T. Padmananbhan,
 Phys. Rep. 380 (2003) 235;
P. J. E. Peebles  and B. Ratra,
Rev. Mod. Phys. 75 (2003) 559;
V. Sahni,
 Class. Quantum Grav. 19 (2002) 3435;
E. J. Copeland, M. Sami and S. Tsujikawa,
Int. J. Mod. Phys. D   15 (2006) 1753;
V. Sahni and A. A. Starobinsky, Int. J. Mod. Phys. D 15 (2006) 2105.
\bibitem{cosmic}
A. Riess et al, 
 Astron. J.  116 (1998) 1009;
S. J. Perlmutter  et al, 
 Astroph. J. 517 (199) 565.
\bibitem{Starobinsky:1998fr}
  A.~A.~Starobinsky,
  JETP Lett.\   68 (1998) 757.
\bibitem{Burd:1988ss}
  A.~B.~Burd and J.~D.~Barrow,
  Nucl.\ Phys.\  B  308 (1988) 929.
\bibitem{Barrow} 
J.~D.~Barrow,
  Phys.\ Lett.\  B  235 (1990) 40.
\bibitem{Gorini:2003wa}
  V.~Gorini, A.~Y.~Kamenshchik, U.~Moschella and V.~Pasquier,
  Phys.\ Rev.\  D  69 (2004) 123512.
\bibitem{Zhuravlev:1998ff}
  V.~M.~Zhuravlev, S.~V.~Chervon and V.~K.~Shchigolev,
  J.\ Exp.\ Theor.\ Phys.\   87 (1998) 223.
\bibitem{Yurov:2003zt}
  A.~Yurov,
  arXiv:astro-ph/0305019.
\bibitem{Yurov:2005bw}
  A.~V.~Yurov and S.~D.~Vereshchagin,
  Theor.\ Math.\ Phys.\  139 (2004) 787.
\bibitem{Guo:2006ab}
  Z.~K.~Guo, N.~Ohta and Y.~Z.~Zhang,
  Mod.\ Phys.\ Lett.\  A  22 (2007) 883.
\bibitem{Guo:2005ata}
  Z.~K.~Guo, N.~Ohta and Y.~Z.~Zhang,
  Phys.\ Rev.\  D 72 (2005) 023504.
\bibitem{Boisseau:2000pr}
  B.~Boisseau, G.~Esposito-Farese, D.~Polarski and A.~A.~Starobinsky,
  Phys.\ Rev.\ Lett.\   85 (2000) 2236.
\bibitem{Andrianov:2007ua}
  A.~A.~Andrianov, F.~Cannata, A.~Y.~Kamenshchik and D.~Regoli,
  JCAP {\bf 0802} (2008) 015
\bibitem{Padmanabhan:2002cp}
  T.~Padmanabhan,
  Phys.\ Rev.\  D 66 (2002) 021301.
\bibitem{Feinstein:2002aj}
  A.~Feinstein,
  Phys.\ Rev.\  D  66 (2002) 063511.
\bibitem{we-Chap}
A.Yu. Kamenshchik, U. Moschella and V. Pasquier, Phys. Lett. B 511 (2001) 265.
\bibitem{modif}
H.~B.~Benaoum,
  Accelerated universe from modified Chaplygin gas and tachyonic fluid, (2002)
  arXiv:hep-th/0205140.
\bibitem{Bilic}
N.~Bilic, G.~B.~Tupper and R.~D.~Viollier,
  Phys.\ Lett.\  B 535 (2002) 17.
\bibitem{we-Chap1}
 V.~Gorini, A.~Kamenshchik and U.~Moschella,
  Phys.\ Rev.\  D  67 (2003) 063509.
\bibitem{we-Chap0}
A.~Kamenshchik, U.~Moschella and V.~Pasquier,
  Phys.\ Lett.\  B  487 (2000) 7.
\bibitem{Bento}
 M.~C.~Bento, O.~Bertolami and A.~A.~Sen,
  Phys.\ Rev.\  D  66 (2002) 043507
\end{thebibliography}
\end{document}